\documentclass[12pt,a4paper]{article}
\usepackage{graphics}
\usepackage{graphicx}
\usepackage{floatrow}
\usepackage{wrapfig}
\usepackage[english]{babel}
\usepackage{mathrsfs}
\usepackage{amsmath,amssymb}
\usepackage{epstopdf}
\usepackage[numbers,sort&compress]{natbib}
\usepackage{braket}
\usepackage{moresize}
\usepackage{tabularx}
\usepackage{color}

\def\FigWidth{\linewidth}

\renewcommand{\vec}{\mathbf}

\begin{document}
\textwidth=135mm
 \textheight=200mm
\begin{center}
{\bfseries Inverse-square law violation and reactor antineutrino anomaly
\footnote{{\small Talk at the International Workshop on Prospects of Particle Physics:
           ``Neutrino Physics and Astrophysics'', Valday, Russia, February 1--8, 2015.
}}}
\vskip 5mm
D.~V.~Naumov, V.~A.~Naumov, D.~S.~Shkirmanov
\vskip 5mm
{\small {\it  Joint Institute for Nuclear Research, 141980 Dubna, Russia}}
\\
\end{center}
\vskip 5mm
\centerline{\bf Abstract}
We discuss a possibility that the  so-called reactor antineutrino anomaly can be, at least in part,
explained by applying a quantum field-theoretical approach to neutrino oscillations, which in particular
predicts a small deviation from the classical inverse-square law at short but macroscopic distances
between the neutrino source and detector. 
An extensive statistical analysis of the reactor data is performed to examine this speculation.
\vskip 10mm

\section{\label{sec:intro} Introduction}

Nuclear reactors are intense sources of electron antineutrinos whose spectrum is composed of thousands
of spectral components formed mainly by the $\beta$ decay of the fission products of the four parent isotopes:
${}^{235}$U, ${}^{238}$U, ${}^{239}$Pu, and ${}^{241}$Pu. 
The very sophisticated recent calculations~\cite{Mueller:2011nm,Huber:2011wv,Mention:2011rk} yield a net 3--3.5\% upward shift
in the predicted energy-averaged $\overline{\nu}_e$ flux with respect to the previously expected flux used in the earlier short
baseline (SBL) reactor experiments 
(ILL--Grenoble        \cite{Kwon:1981ua,Hoummada:1995},
 G\"osgen             \cite{Zacek:1986cu},
 Krasnoyarsk          \cite{Vidyakin:1987ue,Vidyakin:1994ut},
 Rovno                \cite{Afonin:1988gx,Kuvshinnikov:1991,Ketov:1992ee},    
 Bugey                \cite{Declais:1994ma,Achkar:1995},       
 Savannah River Plant \cite{Greenwood:1996pb}).
The $\overline{\nu}_e$ flux normalization uncertainty in the new calculations is claimed to be only $\pm2.7$\%.
This implies~\cite{Mention:2011rk} that the measured event rates in the SBL experiments are about 6\% too low,
giving rise to the so-called ``reactor antineutrino anomaly'' (RAA).

\begin{figure}[htb]
\centering\includegraphics[width=\FigWidth]{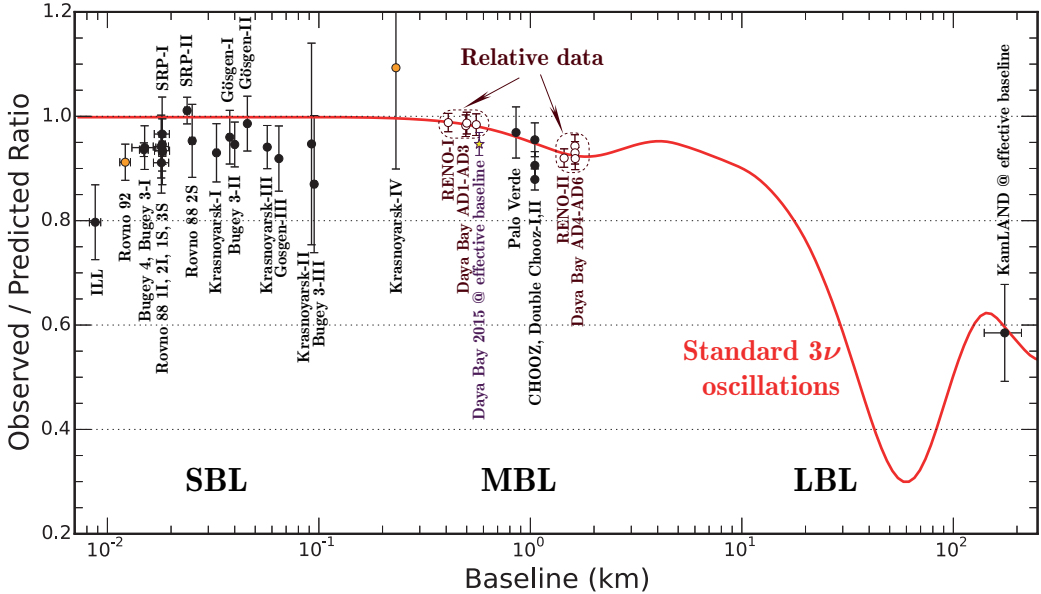}
\protect\caption{\small The reactor antineutrino anomaly. The points represent the ratios of measured event rates to that expected with no oscillations.
                 The vertical error bars do not include the common normalization uncertainty.
                 The original data from Refs.~\cite{Hoummada:1995,Zacek:1986cu,Vidyakin:1987ue,Vidyakin:1994ut,Afonin:1988gx,Kuvshinnikov:1991,Ketov:1992ee,
                 Declais:1994ma,Achkar:1995,Greenwood:1996pb,Boehm:2001ik,Apollonio:2003gd,Abe:2012tg,Abe:2013sxa,Araki:2004mb} are recalculated as
                 explained in the text. The data from Refs.~\cite{Ahn:2012nd,An:2012bu} are relative.
                 The ratios of the KamLAND~\cite{Araki:2004mb} and Daya~Bay~2015~\cite{Naumov:2014pwa} results to the ``Huber-Mueller'' model
                 predictions are plotted at the flux-weighted distances of 175~km and 573~m, respectively.
                 The curve represents the theoretical expectation, in which the $3\nu$ oscillations parameters are fixed to the global best fit values
                 taken from the Review of Particle Physics~\cite{Agashe:2014kda}; the normal mass ordering and no $CP$ violation are assumed.
                }
\label{fig:StandardCurve}
\end{figure}

Figure~\ref{fig:StandardCurve} illustrates this issue.
The curve shows the ratio of the $\overline{\nu}_e$-induced event rate calculated with and without regard for the $3\nu$ oscillations.
Here and thereafter we use the global best fit values for the neutrino mass-squared splittings and mixing angles from Ref.~\cite{Agashe:2014kda}
for the normal mass hierarchy; we also assume no $CP$ violation in mixing. 
The cross section for the inverse $\beta$ decay (IBD) is calculated by using the recent analytical results of Ref.~\cite{Ivanov:2013cga},
which take into account the radiative corrections of order $\alpha/\pi$ and contributions of weak magnetism and neutron recoil
to next-to-leading order in the expansion in inverse powers of the nucleon mass.
In Fig.~\ref{fig:StandardCurve} and below, all the curves correspond to a reactor with pure ${}^{235}$U fuel.%
\footnote{Nevertheless, in the following analysis we explicitly take into account the particular fuel composition in each experiment,
          since it does have a small impact on the results.}
The original data from Refs.~\cite{Hoummada:1995,Zacek:1986cu,Vidyakin:1987ue,Vidyakin:1994ut,Afonin:1988gx,Kuvshinnikov:1991,Ketov:1992ee,
Declais:1994ma,Achkar:1995,Greenwood:1996pb,Boehm:2001ik,Apollonio:2003gd,Abe:2012tg,Abe:2013sxa,Araki:2004mb} are corrected
according to Ref.~\cite{Mueller:2011nm} and then renormalized to the new world average value of the neutron mean life~\cite{Agashe:2014kda}.
It is seen that most of the SBL data points (the measurements at $L\lesssim100$~m, where $L$ is the distance
between the reactor core and detector) are below the expectation. A clear trend is visible at $L\lesssim20$~m that the closer
detector is located to reactor, the smaller the measured rate (the larger the discrepancy between the data and theory). 
Note that the data points ``Krasnoyarsk-IV''~\cite{Vidyakin:1994ut} and ``Rovno\,92''~\cite{Ketov:1992ee} are ignored in the numerous RAA
analyses, but the latter point is significant for revealing the mentioned trend.
Also shown are the data from the medium and long baseline reactor experiments 
Palo~Verde~\cite{Boehm:2001ik},
CHOOZ~\cite{Apollonio:2003gd},     
Double~Chooz~\cite{Abe:2012tg,Abe:2013sxa},
KamLAND~\cite{Araki:2004mb},
RENO~\cite{Ahn:2012nd}, and
Daya~Bay~\cite{An:2012bu,Naumov:2014pwa}.
The data sets from Refs.~\cite{Ahn:2012nd,An:2012bu} are relative measurements, while the recent high-precision
Daya~Bay measurement~\cite{Naumov:2014pwa} is absolute (in Fig.~\ref{fig:StandardCurve} it is placed at the effective baseline of 573~m).
As is seen, the theory is in rather poor agreement with the latter result. Hence, both the earlier SBL and new Daya~Bay measurements
give a hint to either ``new physics'', or merely a lower $\overline{\nu}_e$ flux than predicted in Refs.~\cite{Mueller:2011nm,Huber:2011wv}.

\section{\label{sec:explan} Extra neutrinos or wrong normalization?}

Most if not all efforts to resolve the anomaly are based on the hypothesis of existence of
one or more light (eV mass scale) sterile neutrinos, that is fundamental neutral fermions
with no standard model interactions except those induced by mixing with the standard (active)
neutrinos.
The active-to-sterile neutrino mixing would lead to a distance-dependent spectral distortion
and overall reduction of the reactor $\overline{\nu}_e$ flux.

\begin{figure}[htb]
\centering\includegraphics[width=\FigWidth]{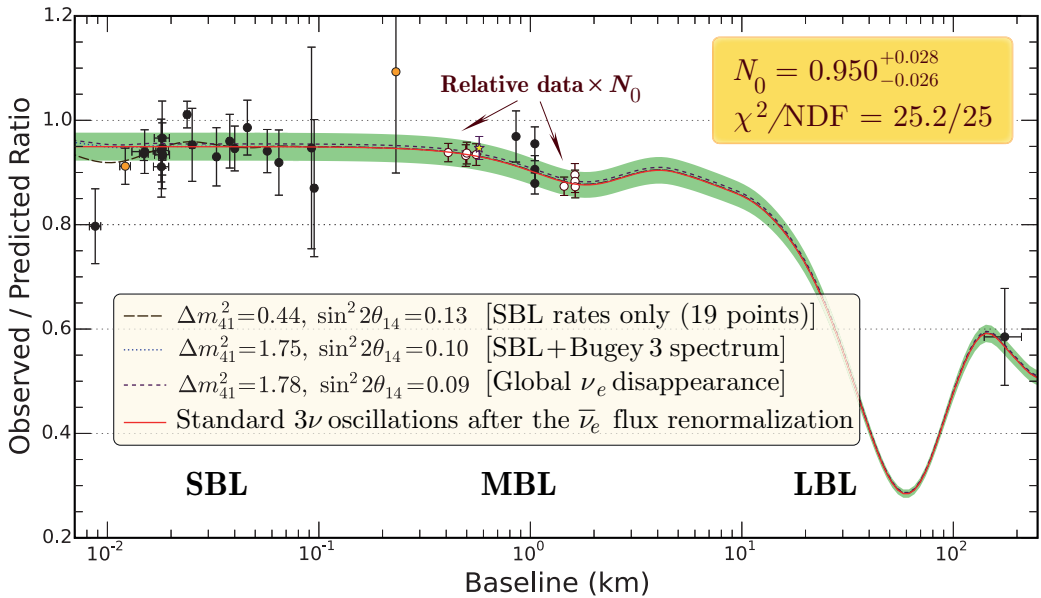}
\protect\caption{\small Comparison of the data with the ``3+1'' model (see text) and with the standard $3\nu$ oscillation
                 prediction after renormalization of the $\overline{\nu}_e$ flux (solid curve).
                 The renormalization factor is obtained from a fit to all the data but RENO and Daya Bay.
                 The filled band shows the $\pm1\sigma$ uncertainty of the fit.
                 The data points are the same as in Fig.~\ref{fig:StandardCurve} but the eight relative data points
                 are shifter by the factor $N_0=0.950$.
                }
\label{fig:N_0}
\end{figure}

In Fig.~\ref{fig:N_0} we show, as an example, the results of calculations performed in the
framework of the simplest ``3+1'' phenomenological model with one sterile (anti)neutrino, $\nu_4$,
by using the three pairs of the $\nu_4-\nu_1$ mixing parameters, $({\Delta}m_{41},\sin^22\theta_{41})$,
listed in the legend of the figure.
These values were derived in Ref.~\cite{Kopp:2013vaa} from detailed statistical analyses
of all the neutrino oscillation data available to date.
The ``SBL rates only'' fit, includes the SBL reactor data except the points ``Krasnoyarsk-IV'' and ``Rovno\,92''
(see Fig.~\ref{fig:StandardCurve}).
The ``SBL + Bugey\,3 spectrum'' fit, includes the same data set and  spectral data from Bugey\,3~\cite{Achkar:1995}.
The ``Global $\nu_e$ disappearance'' fit involves the data from the reactor experiments
\cite{Kwon:1981ua,Zacek:1986cu,Vidyakin:1987ue,Vidyakin:1994ut,Afonin:1988gx,Kuvshinnikov:1991,%
      Declais:1994ma,Achkar:1995,Greenwood:1996pb,Apollonio:2003gd,Boehm:2001ik,Abe:2012tg,Gando:2010aa,Ahn:2012nd,An:2012bu},
as well as solar neutrinos (261 data points from Homestake, SAGE, GALLEX/GNO, Super-Kamiokande, and SNO experiments), 
radioactive source experiments at SAGE and GALLEX,
and the LSND/KARMEN $\nu_e$ disappearance data from $\nu_e-{}^{12}$C scattering 
(see Ref.~\cite{Kopp:2013vaa} for the full list of references and further details). 
It is necessary to note that these fits operate with somewhat lower (to within roughly 1\%)
values for the reactor $\overline{\nu}_e$ induced rates and with a little bit different covariance matrix,
as compared to those used in the present analysis (the details of our calculations will be published elsewhere).

The solid curve in Fig.~\ref{fig:N_0} represents the same $3\nu$ oscillation prediction as in Fig.~\ref{fig:StandardCurve},
but shifted down by the normalization factor $N_0$ derived from a fit to all the data except these from RENO and Daya Bay.
In this fit we take into account the correlation between the data, including the overall normalization uncertainty,
which is taken to be 2.7\%~\cite{Mention:2011rk}. 
The obtained factor $N_0=0.950_{-0.026}^{+0.028}$ ($\chi^2/\text{NDF} \approx 1$) does not contradict to the
adopted flux uncertainty, but is somewhat different from the results of earlier calculations~\cite{Mention:2011rk,Zhang:2013ela,Ivanov:2013cga},
which used different data subsets and input parameters. All four curves in Fig.~\ref{fig:N_0} are in agreement, within the errors,
with the new Day Bay measurement~\cite{Naumov:2014pwa}, but are in some conflict with the ILL data point~\cite{Hoummada:1995}.

Recently, it was argued~\cite{Hayes:2013wra} that the true uncertainty in the $\overline{\nu}_e$ flux predictions may be as large as 5\%
and the spectral shape uncertainties may be much larger due to poorly known structure of the forbidden decays.
This finding has been in essence confirmed by the new precision measurements of the positron energy spectra from IBD~\cite{Naumov:2014pwa},
which show apparent $\sim10$\% excess in $4-6$~MeV region compared to the expectation based on the models of Refs.~\cite{Mueller:2011nm,Huber:2011wv}
(see Ref.~\cite{Dwyer:2014eka} for further discussion and references).
From what has been said it appears that the efforts to explain the anomaly by the sterile neutrino hypothesis may be somewhat premature.
Moreover, it is seen from Fig.~\ref{fig:N_0} that the proper renormalization of the flux is hardly distinguishable from
the ``global $\nu_e$ disappearance'' $4\nu$ fit and (maybe somewhat accidentally) almost fully coincides with the ``SBL + Bugey\,3 spectrum'' fit.
We emphasize however that the steady decrease of the event rate at very small $L$, \emph{if real}, cannot be explained by neither
the wrong $\overline{\nu}_e$ flux normalization alone, nor the ``3+1'' scenario. Thus, it is appropriate to consider an alternative explanation.
Such an alternative has been proposed in Ref.~\cite{Naumov:2013bea}. It is based on a quantum field-theoretical (QFT) approach to
the neutrino oscillation phenomenon, which predicts a small deviation of the (anti)neutrino event rate, as a function of the distance $L$
between the source and detector, from the classical inverse-square law (ISL) behavior. Below we consider this issue in some detail.


\section{A sketch of the QFT approach}

\begin{figure}[htb]
\floatbox[{\capbeside\thisfloatsetup{capbesideposition={right,center},capbesidewidth=0.47\linewidth}}]{figure}[\FBwidth]
{\protect\caption{\small A generic macroscopic Feynman diagram.
                  Here $I_{s,d}$ and $F_{s,d}$ denote the sets of the initial ($I$) and final ($F$) WP states
                  in the ``source'' ($X_s$) and ``detector'' ($X_d$) vertices;
                  $F'_{s,d}=F_{s,d}\oplus\ell_{\alpha,\beta}^{+,-}$, where $\ell_{\alpha,\beta}^{+,-}$ are the
                  charged lepton WP states ($\alpha,\beta=e,\mu,\tau$); $q_{s,d}$ are the 4-momentum transfers in
                  the vertices, as defined by Eq.~\protect\eqref{q_sd}.
                  The vertices are in general \emph{macroscopically} separated in space and time.
                  The particular ``decryption'' of the neutrino production/absorption mechanism assumes the standard model
                  charged current interaction of quarks and leptons.}
\label{fig:Macrograph}}
{\includegraphics[width=\linewidth]{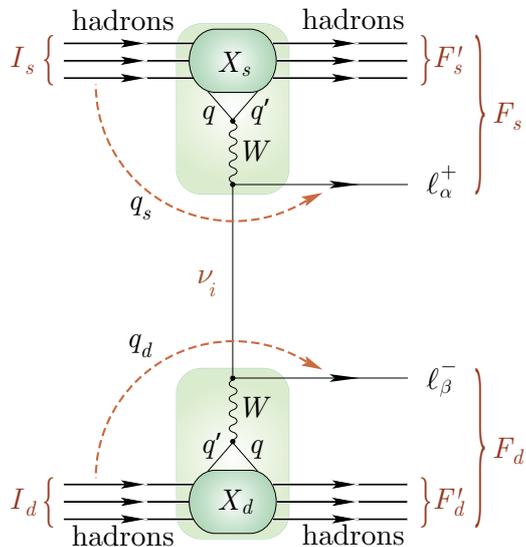}}
\end{figure}

The ``neutrino-oscillation'' phenomenon in the $S$-matrix QFT approach is nothing else than a result of interference
of the macroscopic Feynman diagrams (like shown in Fig.~\ref{fig:Macrograph}) which describe the lepton number
violating processes with the neutrino mass eigenfields $\nu_i$ ($i=1,2,3$) treated as internal lines (propagators).
The external lines of the macrodiagrams are assumed corresponding to asymptotically free quasi-stable wave packets (WP) rather than the conventional
to QFT one-particle Fock's states $|\vec{k},s\rangle$ with definite 3-momenta $\vec{k}$ and spin projections $s$.
According to Refs.~\cite{Naumov:2009zza,Naumov:2010um}, the free external WP states are constructed as covariant space-time point $x$
dependent linear superpositions of the one-particle states,
\begin{equation}
\label{WPdef}
|\vec{p},s,x\rangle = \int\frac{d\vec{k}\,\phi(\vec{k},\vec{p})e^{i(k-p)x}}{(2\pi)^32E_{\vec{k}}}|\vec{k},s\rangle,
\end{equation}
satisfying the \emph{correspondence principle} which demands that $|\vec{p}, s, x\rangle$ turns into $|\vec{k}, s\rangle$
in the plane-wave limit (PWL)
that is equivalent to the following condition for the relativistic invariant form factor function $\phi$:
\[
\phi(\vec{k}, \vec{p}) \stackrel{\text{PWL}}{\longmapsto} (2\pi)^32E_{\vec{p}}\delta(\vec{k}-\vec{p}).
\]
The detail properties of the WP states \eqref{WPdef} are discussed in Refs.~\cite{Naumov:2013vea,Naumov:2014jpa}.

Within the outlined approach, and after applying several  more or less ``technical'' simplifications, it is proved~\cite{Naumov:2010um}
that the neutrino induced event rate in an ideal detector can be written (somewhat symbolically) in the form
\begin{equation}
\label{EventRate_GS}
\frac{dN}{d\tau}
 = \frac{1}{V_{\mathcal{D}}V_{\mathcal{S}}}\int_{V_{\mathcal{S}}} d\vec{x}\int_{V_{\mathcal{D}}} d\vec{y}\int d\mathfrak{F}_\nu
   \int d\sigma_{{\nu}\mathcal{D}}\mathcal{P}_{\alpha\beta}\left(E_\nu,\left|\vec{y}-\vec{x}\right|\right).
\end{equation}
Here $\tau$ is the detector exposure time, $E_\nu$ is the neutrino energy, 
$\mathcal{P}_{\alpha\beta}$ is the QFT generalization of the standard quantum-mechanical neutrino flavor transition probability,
the differential form $d\sigma_{{\nu}\mathcal{D}}$ 
represents the differential cross section of the neutrino scattering from the whole detector device; $d\mathfrak{F}_\nu$
is the differential neutrino flux incident on the detector from a \emph{stationary} source device (e.g., a fission reactor core).
The integrations in \eqref{EventRate_GS} are over the source and detector fiducial volumes $V_{\mathcal{S}}$ and $V_{\mathcal{D}}$. 
The theory explicitly predicts that the neutrino flux decreases with increasing the distance $|\vec{y}-\vec{x}|$ in compliance with
the usual inverse-square law (ISL):
\begin{equation}
\label{TheISL}
d\mathfrak{F}_\nu ~ \propto ~ |\vec{y}-\vec{x}|^{-2}.
\end{equation}
This quite expected result has been derived by using the so-called Grimus-Stockinger (GS) theorem \cite{Grimus:1996av}, which defines
the asymptotic behavior of the amplitude at $L\to\infty$ and this is the crucial point in the context of the problem under consideration.

As it follows from the formalism, the $L$ dependence of the amplitude described by the macrodiagram shown in Fig.~\ref{fig:Macrograph}
is defined by the neutrino propagator modified by the external wave packets,
\begin{equation}
\label{the_propagator}
\left(i\hat{\partial}+m_i\right)\int\frac{d^4{q}}{(2\pi)^4}\;\frac{\widetilde\delta_s(q-q_s)\widetilde\delta_d(q+q_d)e^{-i{qx}}}{{q}^2-m_i^2+i\epsilon},
\end{equation}
where $x=(y_0-x_0,\vec{y}-\vec{x})\equiv(T,\vec{L})$, $q_s$ and $q_d$ are the 4-momentum transfers,
\begin{equation}
\label{q_sd}
q_s = \sum_{a\in{I_s}}p_a-\sum_{b\in{F_s}}p_b,
\quad
q_d = \sum_{a\in{I_d}}p_a-\sum_{b\in{F_d}}p_b,
\end{equation}
$p_\varkappa$ are the most probable (on-shell) 4-momenta of the external packets $\varkappa \in I_s \oplus I_d \oplus F_s \oplus F_d$,
and $m_i$ is the mass of the neutrino field $\nu_i$.
The functions $\widetilde\delta_s(q-q_s)$ and $\widetilde\delta_d(q+q_d)$ are the ``smeared'' $\delta$ functions (see Ref.~\cite{Naumov:2013bea}
for their explicit form) defined by the 4-momenta $p_\varkappa$, masses $m_\varkappa$ ($m_\varkappa^2=p_\varkappa^2$), and momentum spreads
$\sigma_\varkappa$ of the external in and out packets ($\sigma_\varkappa^2 \lll m_\varkappa^2$). 
In the plain-wave limit ($\sigma_\varkappa\to0$, $\forall \varkappa$) these functions turn into the ordinary Dirac $\delta$ functions,
$\widetilde\delta_s(q-q_s)  \stackrel{\text{PWL}}{\longmapsto}  \delta(q-q_s)$,
$\widetilde\delta_d(q+q_d)  \stackrel{\text{PWL}}{\longmapsto}  \delta(q+q_d)$,
thus leading to the exact energy-momentum conservation in the vertices of the macrodiagram,
and the function \eqref{the_propagator} becomes, up to a multiplier, the standard fermion propagator.
If however the momentum spreads $\sigma_\varkappa$ are finite, the space-time behavior of the function \eqref{the_propagator}
is nontrivial. In particular, its spatial dependence at sufficiently large distances $L$ is given by the above-mentioned GS theorem
\cite{Grimus:1996av}, according to which%
\footnote{It is assumed that the complex-valued function $\Phi(\vec{q})$ itself and its first and second derivatives decrease at least like $1/|\vec{q}|^2$
          as $|\vec{q}|\to\infty$ and $\kappa^2>0$; see Ref.~\cite{Naumov:2013bea} for details.}
\[
J(\vec{L},\kappa) = \int\frac{d\vec{q}}{(2\pi)^3}\;\frac{\Phi(\vec{q})e^{i\vec{qL}}}{\vec{q}^2-\kappa^2-i\epsilon}
= \frac{e^{i\kappa L}\Phi(-\kappa\vec{l})}{4{\pi}{L}}\left[1+\mathcal{O}\left(\frac{1}{\sqrt{L}}\right)\right]
\quad
\left(\vec{l}=\frac{\vec{L}}{L}\right)
\]
as $L=|\vec{L}|\to\infty$. This offers the QFT explanation of the ISL behavior \eqref{TheISL} but does not, however, provide the 
spatial scale above which the distance $L$ may be considered as ``sufficiently large''.

In Ref.~\cite{Naumov:2013bea}, an extended version of the GS theorem has been proved, which parametrically defines such a scale
by using the asymptotic expansion of the integral $J(\vec{L},\kappa)$ in terms of inverse powers of $L$ at large $L$.
To be more precise, the theorem in its simplest form states that for any function $\Phi(\vec{q})$ in the Schwartz space $S(\mathbb{R}^3)$
\begin{equation}
\label{TheIntegralExpansion}
J(\vec{L},\kappa) = \frac{e^{i{\kappa}L}}{{4\pi}{L}}\left[\Phi(\vec{q})+\sum_{n \ge 1}\dfrac{(-i)^n
                    D_n\Phi(\vec{q})}{L^n}\right]_{\vec{q}=-\kappa\vec{l}},
\quad
L\to\infty,
\end{equation}
where $D_n$ are explicitly defined differential operators on the momentum space; the lowest order operators, sufficient for our present purpose, are
\begin{align*}
  D_1 = &\  \frac{\kappa}{2}\left[\vec{\nabla}_{\vec{q}}^2-(\vec{l}\vec{\nabla}_{\vec{q}})^2\right]-(\vec{l}\vec{\nabla}_{\vec{q}}),      \\
  D_2 = &\  \frac{\kappa^2}{8}\left[\vec{\nabla}_{\vec{q}}^2-(\vec{l}\vec{\nabla}_{\vec{q}})^2\right]^2 
                 -\kappa(\vec{l}\vec{\nabla}_{\vec{q}})\!\left[\vec{\nabla}_{\vec{q}}^2-(\vec{l}\vec{\nabla}_{\vec{q}})^2\right]
           -\frac{1}{2}\left[\vec{\nabla}_{\vec{q}}^2-3(\vec{l}\vec{\nabla}_{\vec{q}})^2\right].                                          
\end{align*}
Additional important features can be found in Refs.~\cite{Naumov:2013bea,Korenblit:2014uka}.
An analysis of Eq.~\eqref{TheIntegralExpansion} shows that the $1/L$ behavior of the amplitude (and thus the ISL behavior of the event rate)
is violated at the distances $L \lesssim \mathfrak{L}_0$, where
\begin{equation}
\label{TheScale}
\mathfrak{L}_0 \sim \kappa\sigma_{\text{eff}}^{-2} \approx 20\left(\frac{\kappa}{1\;\text{MeV}}\right)
                    \left(\frac{\sigma_{\text{eff}}}{1\;\text{eV}}\right)^{-2}\;\text{cm}
\end{equation}
and the function $\sigma_{\text{eff}}=\sigma_{\text{eff}}(\kappa;\{\vec{v}_\varkappa,m_\varkappa,\sigma_\varkappa\})$ represents
an effective momentum spread dependent on the neutrino momentum $\kappa$ as well as on the mean velocities, masses, and
momentum spreads of the external (in and out) wave packets $\varkappa$. 
The explicit form of this function can be found after specification of a particular model for the external WP states. A simple example
is discussed in Ref.~\cite{Naumov:2013bea} within the so-called contracted relativistic Gaussian packet (CRGP) model~\cite{Naumov:2013vea,Naumov:2014jpa}.
It is in particular shown that $\sigma_{\text{eff}}$ is defined through the transverse (with respect to the neutrino propagation direction $\vec{l}$)
components of the inverse overlap tensors which determine the effective space-time overlap volumes of the WP states in the vertices of the macrodiagram.
It is significant that these components are nearly independent of the neutrino masses (assuming these to be small with respect to the neutrino energy
and thus $\kappa\simeq E_\nu$).
Within the CRGP model, it can be also shown that the magnitude of $\sigma_{\text{eff}}$ is strongly affected by the hierarchy of the external
momentum spreads $\sigma_\varkappa$ but in the simplest case when these spreads are similar in order of magnitude, $\sigma_{\text{eff}}$
is of the same order, too. So, as is seen from Eq.~\eqref{TheScale}, the spatial scale \eqref{TheScale} can be macroscopically large
at sufficiently small external momentum spreads, thus leading to a measurable ISL violation (ISLV).


It is shown in Ref.~\cite{Naumov:2013vea} that Eq.~\eqref{TheIntegralExpansion} modifies the formula for the event rate \eqref{EventRate_GS} 
in such a way that the relation \eqref{TheISL} for the flux is replaced by
\begin{eqnarray}
\label{TheISLviolatinFactor}
d\mathfrak{F}_\nu \propto \frac{1}{|\vec{y}-\vec{x}|^2}
\left(1+{\sum_{n\ge1}\frac{\mathfrak{C}_n}{|\vec{y}-\vec{x}|^{2n}}}\right),
\end{eqnarray}
where the coefficient functions $\mathfrak{C}_n$ are explicitly defined from Eq.~\eqref{TheIntegralExpansion}.
By making expedient assumptions, it can be proved (and this is a crucial point) that $\mathfrak{C}_1<0$.
Hence, using Eq.~\eqref{TheISLviolatinFactor} in leading order (thereby assuming that the ISLV correction is small),
yields the following simple replacement for the event rate:
\begin{equation}
\label{EventRate}
\frac{dN}{d\tau} \longmapsto  \left(1-\frac{\overline{\mathfrak{L}}_0^2}{L^2}\right)\frac{dN}{d\tau}
\end{equation}
(provided that $\overline{\mathfrak{L}}_0^2 \ll L^2$). Here $\overline{\mathfrak{L}}_0\sim\langle\mathfrak{L}_0\rangle$ is a neutrino energy
dependent parameter of dimension of length. Needless to say, at present this parameter cannot be obtained from first-principle calculations,
but it can be measured.


\section{Data analysis}

\begin{figure}[h]
\centering\includegraphics[width=\FigWidth]{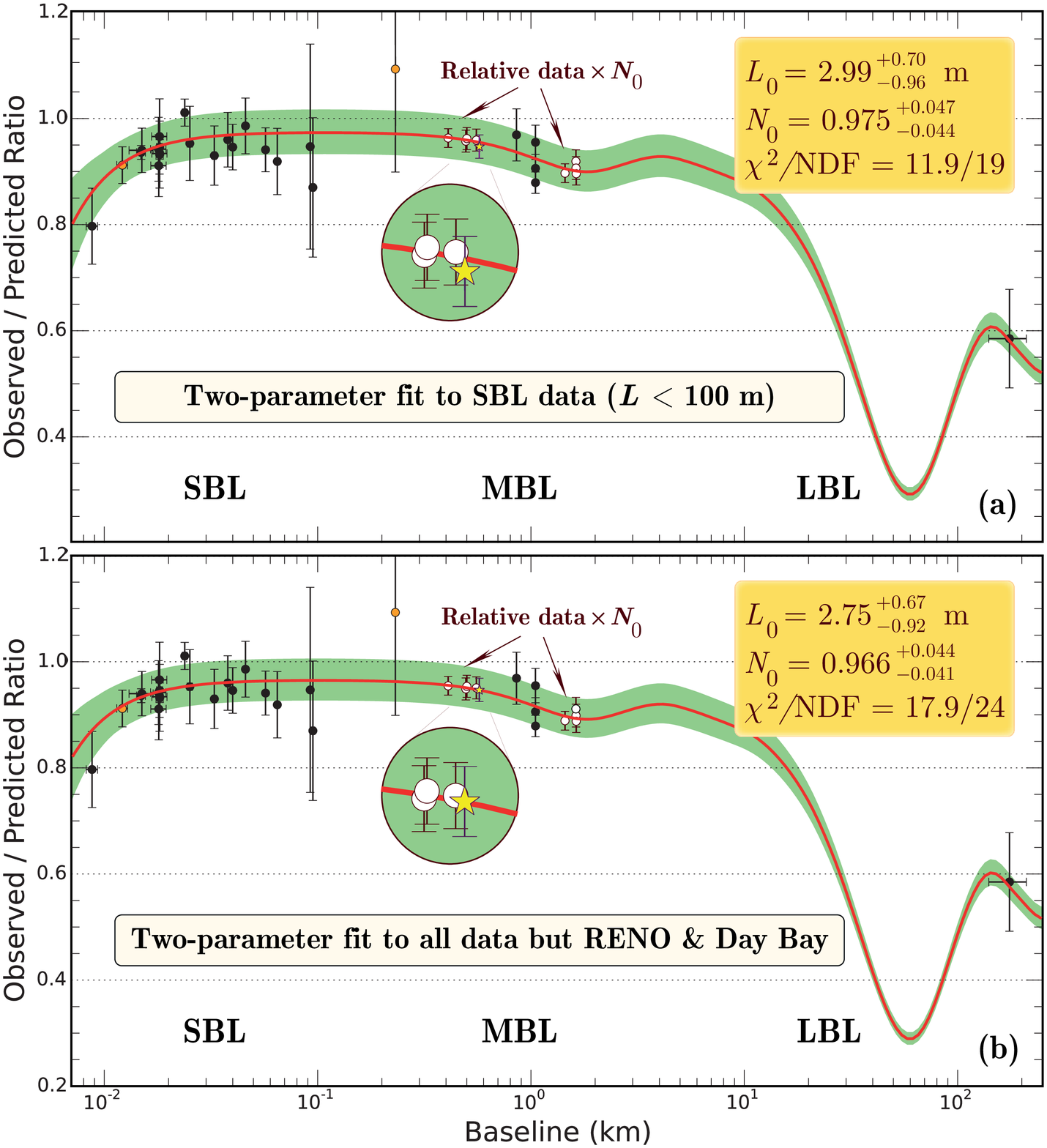}
\protect\caption{\small Two-parameter fits to SBL data subset (a) and to all data (b).
                 The data of RENO and Daya Bay are not included into the analysis.
                 The eight relative data points in the panels (a) and (b) are shifter by the factor
                 $N_0=0.975$ and $0.966$, respectively. 
                 The inserts clarifies the agreement between the best-fit curves and Daya Bay data points.}
\label{fig:L0N0_2_panels}
\end{figure}

\begin{figure}[htb]
\centering\includegraphics[width=\textwidth]{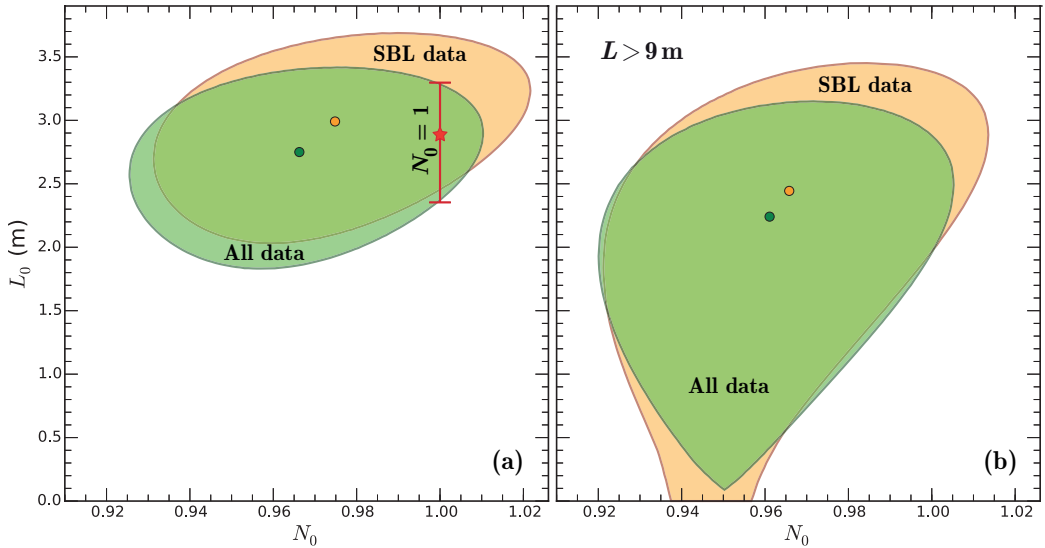}
\protect\caption{\small Panel (a): The $68$\% C.L.\ error contours for the pair of the parameters $N_0$ and $L_0$
                        extracted from the fits to all data and SBL data subset ($L<300$~m). 
                        The one-parameter (with $N_0$ fixed to 1) fit to SBL data is also shown for comparison.
                        Panel (b): The same as in (a), but with the ILL data point excluded from the statistical analysis.
                        In both panels, the circles correspond to the best-fit values of the fitted parameters
                        with $L_0^{\text{SBL data}}>L_0^{\text{All data}}$ and $N_0^{\text{SBL data}}>N_0^{\text{All data}}$.
                }
\label{fig:Contours}
\end{figure}

To check the assumption that the ISLV effect could actually be, in part, responsible for RAA, we performed
a statistical analysis of the reactor data discussed in Sect.~\ref{sec:intro}. Since in this paper we use 
only the spectrum-averaged event rates, the $L$ dependent factor in Eq.~\eqref{EventRate} can be replaced
by $(1-L_0^2/L^2)$, where $L_0$ $\sim\langle\overline{\mathfrak{L}}_0\rangle$ is an energy independent parameter,
which is a subject of the present study. Taking into account the large uncertainty in the $\overline{\nu}_e$ flux
normalization, we shall use the following theoretical model to fit the data:
\begin{equation}
\label{eq:fit_prediction}
T(L;N_0,L_0) = 
N_0\left(1-\frac{L_0^2}{L^2}\right)
\dfrac{\int_0^\infty dE_{\nu} \sum_k f_k P^{3\nu}_\text{surv}(L,E_{\nu})\sigma(E_{\nu})
S_k(E_{\nu})}{\int_0^\infty dE_{\nu} \sum_k f_k \sigma(E_{\nu}) S_k(E_{\nu})}.
\end{equation}
Here $N_0$ is the required normalization parameter, 
$f_k$ is the reactor fissile isotope fraction, 
$S_k(E_{\nu})$ is the $\overline{\nu}_e$ energy spectrum (taken from Ref.~\cite{Mueller:2011nm}), 
$\sigma(E_{\nu})$ is the IBD cross section~\cite{Ivanov:2013cga} and
$P^{3\nu}_\text{surv}(L,E_{\nu})$ is the $\overline{\nu}_e$ survival probability in the standard $3\nu$ mixing scheme.%
\footnote{We thereby neglect the decoherence effects predicted in the QFT approach~\cite{Naumov:2010um}, reasonably assuming that the baselines
          under consideration are too short for their manifestation.}
In order to find the best-fit parameters $N_0$ and $L_0$ we minimize the standard $\chi^2$ with the full covariance matrix for the correlated data.
\begin{figure}[hb]
\centering\includegraphics[width=\FigWidth]{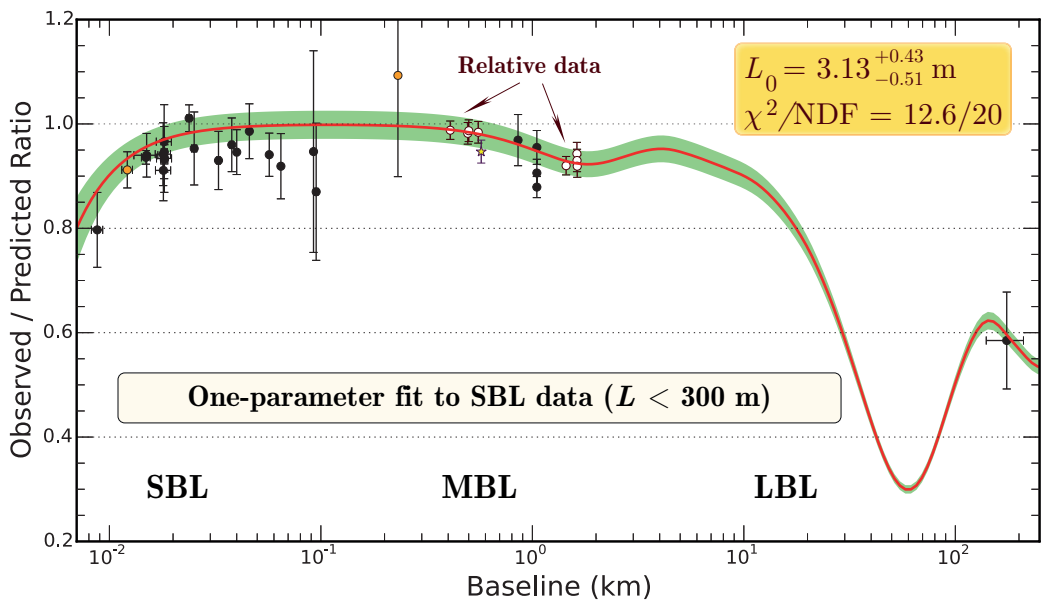}
\protect\caption{\small One-parameter fit to SBL data assuming $N_0=1$.}
\label{fig:L_0_SBL}
\end{figure}
The results of the fits of several types are shown in Figs.~\ref{fig:L_0_SBL}--\ref{fig:Contours}. 

\clearpage

\clearpage

Figure~\ref{fig:L_0_SBL} represents the results of the simplest one-parameter fit, in which $N_0$ is set to 1 and only the SBL data ($L<300$~m)
are used in the analysis. 
Here and below, the filled band represents the $\pm1\sigma$ uncertainty.
It is seen that despite relatively small value of $\chi^2/\text{NDF}$ the best-fit curve is in rather poor agreement with the data;
in particular, it is in conflict with the recent Daya Bay measurement. This indicates that the ISLV suppression alone is not
sufficient and the flux renormalization is actually required.

Figure \ref{fig:L0N0_2_panels} illustrates the results of the two-parameter fits performed with two different data subsets, 
namely with the SBL data only, and with all the data except these from RENO and Daya Bay. As is seen from the figure,
these two fits are in reasonable agreement to each other and both describe the reactor data rather well.
Although the value of $\chi^2/\text{NDF}$ for the SBL data is nominally a bit better than that for the fit to all data
(0.63 and 0.74, respectively), the latter fit is (as is clearly seen in the inserts of Fig.~\ref{fig:L0N0_2_panels}) in
better agreement with the Daya Bay point (let us remind that it does not participate in the analysis).
By comparing these results with the fits shown in Fig.~\ref{fig:N_0}, we may conclude that the ISLV effect in combination
with the proper renormalization of the $\overline{\nu}_e$ flux provides a better resolution of the anomaly.
 
To gain a deeper understanding of our results, we compare in Fig.~\ref{fig:Contours} the $68$\% C.L.\ error contours for the
pair of the fitted parameters $(N_0,L_0)$, obtained from the fits to different data subsets. Panel (a) in Fig.~\ref{fig:Contours}
shows the contours for the fits to all data and SBL data subset. Panel (b) shows the same but with the ILL data point excluded
from the analysis. It is seen with no need of additional explanation that the Grenoble experiment is the ``cornerstone'' for
verification of the ISLV effect. It is however important that even without the ILL point we obtain essentially the same best-fit
values of the parameters. 

\section{Conclusions}

The QFT approach predicts a deviation from the classical inverse-square law at short baselines.
While the numerical value of the spatial scale at which the deviation becomes essential cannot
be predicted from the present-day theory, it can be extracted, under reasonable assumptions,
from the data of the past and current reactor antineutrino experiments. Our statistical analysis
demonstrates that the averaged over the reactor antineutrino spectrum value of the scale ($L_0$)
is about 3~m that roughly corresponds to the spectrum-averaged effective momentum spread
$\langle\sigma_{\text{eff}}\rangle$ of about $0.5-0.8$~eV (thereby hinting that the
wave packets of the particles and nuclei involved into the reactor $\overline{\nu}_e$
production and detection may have ``mesoscopic'' effective dimensions).
This is in agreement with the conservative estimate presented in Ref.~\cite{Naumov:2013bea}.
Besides, the best-fit value of $L_0$ is very stable with respect to choice of the data subset
and $\overline{\nu}_e$ spectrum model. To check the latter, we performed the same one- and
two-parameter fits as described above, but with the input $\overline{\nu}_e$ energy spectra
from Refs.~\cite{Fallot:2012jv} and \cite{Sinev:2013ova} derived by very different methods,
as well as with combinations of the models \cite{Mueller:2011nm,Huber:2011wv} and
the cumulative $\overline{\nu}_e$ spectrum from ${}^{238}$U fission recently measured with
the scientific neutron source FRM\,II in Garching \cite{Haag:2013raa}.
We conclude from these exercises that value of $L_0$ is almost insensitive (within the errors)
to the spectrum variations.
Needless to say, it is not the case for the normalization parameter $N_0$.

Although the available reactor data cannot definitely confirm or exclude the light sterile
neutrino hypothesis, and do not provide unambiguous support for the ISLV effect, they are
in much better agreement with the latter. 
The next-generation experiments with very short baselines ($L\lesssim 20$~m), small neutrino
or antineutrino sources, and high-precision, desirably movable detectors are required in order
to confirm or disconfirm our explanation.

{\small
}

\end{document}